\documentclass[aps,prd,twocolumn,superscriptaddress,showpacs,showkeys]{revtex4-2}
\usepackage{color,xcolor}
\usepackage{amssymb}
\usepackage{graphicx}
\usepackage{extarrows}

\newcommand{\bastar}{\begin{eqnarray*}}
\newcommand{\eastar}{\end{eqnarray*}}
\newskip\humongous \humongous=0pt plus 1000pt minus 1000pt

\newif\ifdtup

\relax
\newcommand{\W}{{\vec W}}
\newcommand{\n}{\hat n}

\newcommand{\hr}{\hat r}

\newcommand{\hD}{{\hat D}}

\newcommand{\bea}{\begin{eqnarray}}
\newcommand{\eea}{\end{eqnarray}}
\newcommand{\pd}{\partial}

\newcommand{\Int}{\displaystyle\int}
\newcommand{\A}{{\vec A}}
\newcommand{\hA}{{\hat A}}
\newcommand{\B}{{\vec B}}

\newcommand{\tA}{{\tilde A}}
\newcommand{\tC}{{\tilde C}}
\newcommand{\F}{{\vec F}}

\newcommand{\hF}{{\hat F}}

\newcommand{\bg}{\bar g}
\newcommand{\mn}{{\mu\nu}}
\newcommand{\al}{\alpha}

\newcommand{\vsig}{{\vec \sigma}}
\newcommand{\vp}{{\varphi}}
\newcommand{\nn}{\nonumber}
\newcommand{\pro}{\partial}

\newcommand{\eps}{{\epsilon}}

\newcommand{\beps}{{\bar \epsilon}}
\begin{document}
\title{Electroweak Strings in the Standard Model}

\author{Li-Ping Zou}
\email{zoulp5@mail.sysu.edu.cn}
\affiliation{Sino-French Institute of Nuclear Engineering and 
Technology, Sun Yat-Sen University, Zhuhai 519082, China}
\author{Pengming Zhang}
\email{zhangpm5@mail.sysu.edu.cn}
\affiliation{School of Physics and Astronomy,
Sun Yat-Sen University, Zhuhai 519082, China}
\author{Y. M. Cho}
\email{ymcho0416@gmail.com}
\affiliation{School of Physics and Astronomy,
Seoul National University, Seoul 08826, Korea}
\affiliation{Center for Quantum Spacetime,
Sogang University, Seoul 04107, Korea}

\begin{abstract}
We argue that the existence of the electroweak monopole
predicts the existence of the electroweak string in 
the standard model made of monopole-antimonopole pair 
separated infinitely apart, which carry the quantized 
magnetic flux $4 \pi n/e$. We show how to construct such quantized magnetic flux string solution. Our result 
strongly indicates that genuine fundamental electromagnetic 
string could exist in nature which could actually be 
detected. We discuss the physical implications of 
our result in cosmology.
\end{abstract}
\maketitle

\section{Introduction}

With the recent discovery of Higgs particle at LHC, it has
been widely regarded that the standard model has passed
the ``final" test \cite{LHC}. This has urged people to go
``beyond" the standard model. But we emphasize that this
view might be premature, because it has yet to pass another
important test, the topological test. By now it is well
known that the standard model must have the electroweak
(Cho-Maison) monopole as the electroweak generalization 
of Dirac monopole \cite{plb97,yang}. And this is within, 
not beyond, the standard model. This means that the true 
final test of the standard model should come from 
the discovery of the topological objects of the model, 
in particular the electroweak monopole.

After Dirac predicted the existence of the monopole,
the monopole has become an obsession \cite{dirac,cab}.
But the Dirac monopole, in the course of the electroweak
unification of the weak and electromagnetic interactions,
changes to the electroweak monopole \cite{plb97,yang}. 
So, the monopole which should exist in the real world 
is not the Dirac monopole but this one. This has 
triggered new studies on the electroweak monopole \cite{epjc15,ellis,ak,bb,mav,pta19,epjc20,fh,gv1,gv2}.
Moreover, if detected it will become the first magnetically charged topological elementary particle in the history of 
physics. For this reason MoEDAL and ATLAS at LHC, IceCube 
at the south pole as well as other detectors are actively searching for such monopole \cite{medal,atlas,ice}.  

If the standard model has the monopole, it must also
have another important topological object, the quantized
magnetic vortex. This is because the electroweak monopole
predicts the existence of the electroweak string made of
the monopole-antimonopole pair which carries the quantized
magnetic flux $4\pi n/e$. This is not surprising. Since 
the standard model includes the unbroken electromagnetic 
U(1) interaction it is natural to expect that it has 
the magnetic vortex. {\it The purpose of this paper is 
to demonstrate the existence of the electromagnetic string 
carrying the quantized magnetic flux in the standard model, and to discuss the physical implications of the string.}

Of course, it has been well known that the standard model
has the string solution known as the Nambu string, W string, 
and Z-string \cite{nambu,vacha,bvb}. But here we are talking 
about a new type of string, the electromagnetic string 
which carries quantized magnetic flux.

The existence of the electromagnetic string in the standard
model should have deep implications in physics. Clearly
it could be interpreted as a fundamental topological string
in nature, so that it has an important theoretical meaning.
Moreover, it could play an important role in cosmology,
in particular in the early universe in the formation of 
large scale structure of the universe \cite{witt}.

The paper is organized as follows. In Section II we review
the Abelian decomposition of the standard model for to
clarify the topological structure of the model. In Section 
III we review the electroweak monopole for later use. In
Section IV we discuss the possible string ansatz and string equation of motion in the standard model, in particular 
the magnetic vortex solution made of the infinitely
separated monopole-antimonopole pair. In Section V we discuss 
the electroweak vortex solution carrying quantized magnetic 
flux in the standard model. In Section VI we compare this 
string solution with the known Z-string solution. Finally 
in the last section we discuss the physical implications 
of our results.

\section{Abelian Decomposition of the Standard Model: A Review}

Before we discuss the electroweak string it is important 
for us to understand the structure of the standard model. 
For this we start from the gauge independent Abelian
decomposition of the electroweak theory. Consider 
the (bosonic sector of) Weinberg-Salam model,
\begin{gather}
{\cal L} =-|{\cal D}_\mu \phi|^2
-\frac{\lambda}{2}\big(|\phi|^2
-\frac{\mu^2}{\lambda}\big)^2-\frac{1}{4}\F_\mn^2
-\frac{1}{4}G_\mn^2, \nn \\
{\cal D}_\mu \phi =\big(\pd_\mu
-i\frac{g}{2} \vsig \cdot \A_\mu
-i\frac{g'}{2} B_\mu\big) \phi  \nn\\
=D_\mu \phi-i\frac{g'}{2} B_\mu \phi,
\label{lag0}
\end{gather}
where $\phi$ is the Higgs doublet, $\A_\mu$, $\F_\mn$
and $B_\mu$, $G_\mn$ are the gauge fields of $SU(2)$ 
and hypercharge $U(1)_Y$, and $D_\mu$ is the covariant 
derivative of $SU(2)$.  With
\begin{gather}
\phi = \dfrac{1}{\sqrt{2}} \rho~\xi,~~~(\xi^\dagger \xi = 1),
\end{gather}
we have
\begin{gather}
{\cal L}=-\frac{1}{2} (\pd_\mu \rho)^2
- \frac{\rho^2}{2} |{\cal D}_\mu \xi |^2
-\frac{\lambda}{8}\big(\rho^2-\rho_0^2 \big)^2 \nn\\
-\frac14 \F_\mn^2 -\frac14 G_\mn^2,
\label{lag1}
\end{gather}
where $\rho_0=\sqrt{2\mu^2/\lambda}$ is the vacuum value
of the Higgs field. Notice that the $U(1)_Y$ coupling
of $\xi$ makes the theory a gauge theory of $CP^1$
field \cite{plb97}.

Let $(\n_1,\n_2,\n_3=\n)$ be an arbitrary right-handed
orthonormal basis of $SU(2)$. Identifying $\n$ to be
the Abelian direction at each space-time point, we have
the Abelian decomposition of the $SU(2)$ gauge field to
the restricted part $\hA_\mu$ and the valence part
$\W_\mu$ \cite{prd80,prl81},
\begin{gather}
\A_\mu = \hA_\mu + \W_\mu,     \nn\\
\hA_\mu =\tA_\mu +\tC_\mu,
~~~\W_\mu =W^1_\mu ~\n_1 + W^2_\mu ~\n_2, \nn\\
\tA_\mu= A_\mu \n~~(A_\mu=\n \cdot \A_\mu),
~~~\tC_\mu=-\frac{1}{g} \n\times \pd_\mu \n,  \nn\\
\F_\mn=\hF_\mn + \hD _\mu \W_\nu - \hD_\nu
\W_\mu + g\W_\mu \times \W_\nu,   \nn\\
\hD_\mu=\pd_\mu+g \hA_\mu \times.
\label{cdec}
\end{gather}
It has the following properties. First, $\hA_\mu$ is
made of two potentials, the non-topological $\tA_\mu$
and topological $\tC_\mu$. Second, it has the full $SU(2)$
gauge degrees of freedom, in spite of the fact that
it is restricted. Third, $\W_\mu$ transforms gauge
covariantly. Most importantly, the decomposition is
gauge independent. Once the Abelian direction is chosen
it follows automatically, regardless of the choice of
gauge.

The restricted field strength $\hF_\mn$ inherits
the dual structure of $\hA_\mu$, which can also
be described by two potentials $A_\mu$ and $C_\mu$,
\begin{gather}
\hF_\mn= \pd_\mu \hA_\nu-\pd_\nu \hA_\mu
+ g \hA_\mu \times \hA_\nu =F_\mn' \n, \nn \\
F'_\mn=F_\mn + H_\mn
= \pd_\mu A'_\nu-\pd_\nu A'_\mu,  \nn\\
F_\mn =\pd_\mu A_\nu-\pd_\nu A_\mu, \nn\\
H_\mn = -\frac1g \n \cdot (\pd_\mu \n \times\pd_\nu \n)
=\pd_\mu C_\nu-\pd_\nu C_\mu,  \nn\\
\n=-\xi^\dagger \vsig \xi,   \nn\\
C_\mu \simeq -\frac1g \n_1\cdot \pd_\mu \n_2
\simeq -\frac{2i}{g} \xi^\dagger \pd_\mu \xi,   \nn\\
A_\mu' = A_\mu+ C_\mu.
\end{gather}
Although $\hF_\mn$ has only the Abelian component,
it describes the non-trivial U(1) gauge theory, because
it has not only the Maxwellian $A_\mu$ but also 
the Diracian $C_\mu$ which describes the monopole 
potential \cite{prd80,prl81}. This justifies
us to call $A_\mu$ and $C_\mu$ the non-topological
electric and topological magnetic potential.

With the Abelian decomposition we can express (\ref{lag0})
in terms of physical fields in an Abelian form gauge
independently. To do this we first define the electromagnetic
and Z boson fields $A_\mu^{\rm (em)}$ and $Z_\mu$ with
the Weinberg angle by
\begin{gather}
\left(\begin{array}{cc} A_\mu^{\rm (em)} \\  Z_{\mu}
\end{array} \right)
=\frac{1}{\sqrt{g^2 + g'^2}} \left(\begin{array}{cc} g & g' \\
-g' & g \end{array} \right)
\left(\begin{array}{cc} B_{\mu} \\ A'_{\mu}
\end{array} \right)  \nn\\
= \left(\begin{array}{cc}
\cos\theta_{\rm w} & \sin\theta_{\rm w} \\
-\sin\theta_{\rm w} & \cos\theta_{\rm w}
\end{array} \right)
\left( \begin{array}{cc} B_{\mu} \\ A_\mu'
\end{array} \right).
\label{mix}
\end{gather}
Now, with the identity
\begin{gather}
D_\mu \xi= -i\frac{g}{2} \big[(A'_\mu \n
+\W_\mu) \cdot \vsig\big]~\xi,  \nn\\
|D_\mu \xi|^2 =\frac{g^2}{4} ({A'}_\mu^2
+ \W_\mu^2),  \nn\\
|{\cal D}_\mu \xi|^2 =\frac{g^2+g'^2}{8} Z_\mu^2
+\frac{g^2}{4} \W_\mu^2,
\label{id1}
\end{gather}
we can ``abelianize" (\ref{lag0}) to
\begin{gather}
{\cal L}= -\frac12 (\pd_\mu \rho)^2
-\frac{\lambda}{8}\big(\rho^2-\rho_0^2 \big)^2
-\frac14 {F_\mn^{\rm (em)}}^2   \nn\\
-\frac12 \big|(D_\mu^{\rm (em)} +ie\frac{g}{g'} Z_\mu) W_\nu
-(D_\nu^{\rm (em)} +ie\frac{g}{g'} Z_\nu) W_\mu \big|^2  \nn\\
-\frac14 Z_\mn^2-\frac{\rho^2}{4} \big(g^2 W_\mu^*W_\mu
+\frac{g^2+g'^2}{2} Z_\mu^2 \big)   \nn\\
+ie (F_\mn^{\rm (em)}
+ \frac{g}{g'}  Z_\mn) W_\mu^* W_\nu  
+ \frac{g^2}{4}(W_\mu^* W_\nu - W_\nu^* W_\mu)^2,  \nn\\
D_\mu^{\rm (em)}=\pd_\mu+ieA_\mu^{\rm (em)},
~~~W_\mu =\frac{1}{\sqrt 2} (W^1_\mu + i W^2_\mu),   \nn\\
e=\frac{gg'}{\sqrt{g^2+g'^2}}=g\sin\theta_{\rm w}
=g'\cos\theta_{\rm w}.
\label{lag2}
\end{gather}
This is the gauge independent Abelianization of the standard 
model.

From (\ref{lag2}) we obtain the following equations
of motion
\begin{gather}
\pd^2 \rho-\big(\frac{g^2}{2} W_\mu^*W_\mu
+\frac{g^2+g'^2}{4}Z_\mu^2 \big)~\rho
=\frac{\lambda}{2}\big (\rho^2 -\rho_0^2 \big)~\rho,   \nn\\
\pd_\mu \Big[F_\mn^{\rm(em)}
-ie(W_\mu^* W_\nu-W_\nu^* W_\mu) \Big]
=ie \Big[W_\mu^* (D_\mu^{\rm (em)} W_\nu   \nn\\
-D_\nu^{\rm (em)} W_\mu) -(D_\mu^{\rm (em)} W_\nu
-D_\nu^{\rm (em)} W_\mu)^* W_\mu \Big]  \nn\\
+ e^2 \frac{g}{g'} \Big[2 W_\mu^*W_\mu Z_\nu-Z_\mu(W_\mu^*W_\nu+W_\nu^*W_\mu) \Big],  \nn\\
\Big(D_\mu^{\rm (em)}+ie\frac{g}{g'}Z_\mu\Big) 
\Big[D_\mu^{\rm (em)} W_\nu -D_\nu^{\rm (em)} W_\mu  \nn\\
+ie\frac{g}{g'}(Z_\mu W_\nu-Z_\nu W_\mu)\Big]
=\frac{g^2}{4}\rho^2 W_\nu  \nn\\
+ i e W_\mu \big(F^{\rm(em)}_\mn + \frac{g}{g'} Z_\mn \big)
+ g^2 W_\mu(W_\mu^* W_\nu-W_\nu^* W_\mu),\nn\\
\pd_\mu \Big[Z_\mn-ie\frac{g}{g'} (W_\mu^* W_\nu
- W_\mu W_\nu^*) \Big] 
-\frac{g^2+g'^2}{4} \rho^2 Z_\nu  \nn\\
=ie\frac{g}{g'}\Big[W_\mu^*(D_\mu^{\rm (em)} W_\nu
-D_\nu^{\rm (em)} W_\mu)   \nn\\
-W_\mu (D_\mu^{\rm (em)} W_\nu
-D_\nu^{\rm (em)} W_\mu) \Big]  \nn\\
+e^2\frac{g^2}{g'^2}\Big[ 2W_\mu^*W_\mu Z_\nu-Z_\mu(W_\mu^*W_\nu+W_\nu^*W_\mu)\Big].
\label{wseq2}
\end{gather}
This should be compared with the equations of
motion obtained from (\ref{lag1}),
\begin{gather}
\pro^2\rho =|{\cal D}_\mu \xi |^2 \rho
+\frac{\lambda}{2}\big (\rho^2-\rho_0^2 \big) \rho, \nn\\
{\cal D}^2 \xi=-2 \dfrac{\pd_\mu \rho}{\rho}
{\cal D}_\mu \xi +\big[\xi^\dagger {\cal D}^2\xi
+2\dfrac{\pd_\mu \rho}{\rho}
(\xi^\dagger {\cal D}_\mu \xi) \big] \xi, \nn\\
D_\mu \F_\mn=i \frac{g}{4}\rho^2 \big[\xi^\dagger
\vec \tau( {\cal D}_\nu \xi )
-({\cal D}_\nu \xi)^\dagger \vec \tau \xi \big], \nn\\
\pd_\mu G_\mn
=i\frac{g'}{4}\rho^2 \big[\xi^\dagger ({\cal D}_\nu \xi)
- ({\cal D}_\nu \xi)^\dagger \xi \big].
\label{em0}
\end{gather}
The contrast is unmistakable.

This Abelianization teaches us an important lesson,
the assertion that the Higgs mechanism comes from 
the spontaneous symmetry breaking, is a simple 
misunderstanding. To see this notice that (\ref{lag2}) 
is mathematically identical to (\ref{lag0}). This means 
that (\ref{lag2}) still retains the full non-Abelian 
$SU(2)\times U(1)_Y$ gauge symmetry. In fact, the non-Abelian gauge symmetry is hidden in the fact that $A_\mu^{(em)}$ 
and $Z_\mu$ are defined in terms of $A'_\mu$ made of two potentials $A_\mu$ and $C_\mu$. 

Nevertheless, the W and Z bosons acquire mass when $\rho$ 
has the non-vanishing vacuum value in (\ref{lag2}). This 
means that we have the Higgs mechanism (i.e., the mass 
generation) without any (spontaneous or not) symmetry 
breaking. In fact, in (\ref{lag2}) we have no Higgs doublet 
which can break the $SU(2)\times U(1)_Y$ symmetry. This 
confirms that the Higgs mechanism has nothing to do with 
the spontaneous symmetry breaking.  

The above exercise tells us that the standard model has 
another important topology, the $\pi_1(S^1)$ string 
topology which implies the existence of a topological 
string, in particular the electromagnetic string. Actually 
the Abelian decomposition tells that the standard model 
has two $\pi_1(S^1)$ topology, coming from the unbroken electromagnetic $U(1)$ and the spontaneously broken Z boson 
$U(1)$, which strongly implies the existence of two types 
of strings.

Indeed, in the absence of weak bosons (\ref{lag2})
reduces to the non-trivial electrodynamics
\begin{gather}
{\cal L}=-\frac14 {F_\mn^{\rm (em)}}^2,
\label{lag4}
\end{gather}
which has the Dirac-type electroweak monopole of
magnetic charge $4\pi/e$. This is because the electromagnetic 
$U(1)_{em}$ here is non-trivial which has the period 
$4\pi$ (not $2\pi$). This is evident from (\ref{mix}),
which tells that $U(1)_{em}$ comes from the superposition 
of the $U(1)$ subgroup of $SU(2)$. 

This strongly implies that it also admits the electromagnetic string made of the monopole-antimonopole pair carrying 
the magnetic flux $4\pi/e$. Moreover, in the absence of $A_\mu^{\rm (em)}$ and $W$-boson (\ref{lag2}) reduces to 
the spontaneously broken $U(1)_Z$ gauge theory,
\begin{gather}
{\cal L}=-\frac12 (\pd_\mu \rho)^2
-\frac{\lambda}{8}(\rho^2-\rho_0^2)^2 -\frac14 Z_\mn^2 \nn\\
-\dfrac{g^2+g'^2}{8} \rho^2 Z_\mu^2,
\label{lag3}
\end{gather}
which describes the Landau-Ginzburg Lagrangian of 
superconductivity which is well known to have 
the Abrokosov-Nielsen-Olesen (ANO) vortex
solution \cite{ano}. In the following we show how 
to obtain such string solutions in the standard model.

\section{Electroweak Monopole: A Review}

Before we discuss the string solution, we review
the electroweak monopole, because the string
is deeply related to the monopole-antimonopole pair.
Start from the monopole ansatz in the spherical
coordinates $(t,r,\theta,\varphi)$ \cite{plb97,yang}
\begin{gather}
\phi=\frac{i}{\sqrt{2}} \rho(r)
\left(\begin{array}{cc}
\sin \dfrac{\theta}{2}~\exp(-i\varphi) \\
- \cos \dfrac{\theta}{2} \end{array} \right), \nn\\
\A_\mu= \frac{1}{g} A(r) \pd_\mu t~\hr
+\frac{1}{g}(f(r)-1)~\hr \times \pd_\mu \hr, \nn\\
B_{\mu} =\frac{1}{g'} B(r) \pd_\mu t
-\frac{1}{g'}(1-\cos\theta) \pd_\mu \varphi.
\label{ans0}
\end{gather}
In terms of physical fields the ansatz becomes
\begin{gather}
\rho =\rho(r),  \nn\\
W_\mu= \frac{i}{g} \frac{f}{\sqrt 2}e^{i\varphi}
(\pd_\mu \theta +i \sin\theta \pd_\mu \varphi), \nn\\
A_\mu^{\rm (em)}=e\big(\frac{A}{g^2}+\frac{B}{g'^2} \big)
\pd_\mu t -\frac1e (1-\cos\theta)\pd_\mu \varphi,  \nn\\
Z_\mu= \frac{e}{gg'}\big(A-B \big)\pd_\mu t.
\label{ans1}
\end{gather}
This clearly shows that the ansatz is for the electroweak
dyon.

\begin{figure}
\includegraphics[height=4.5cm, width=8cm]{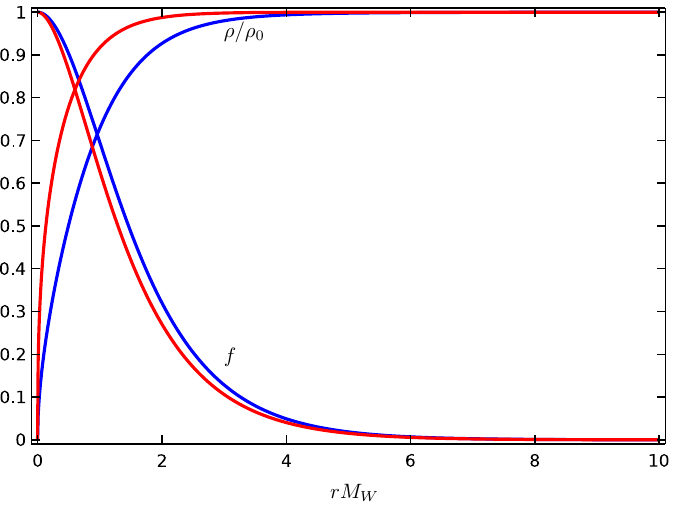}
\caption{\label{fem} The electroweak monopole solution.
The red and blue curves represent the singular Cho-Maison 
monopole and the regularized finite energy monopole 
obtained by (\ref{fedeq1}) with $A=B=0$ and $n=6$.}
\end{figure}

With the ansatz we have the following equations of motion
\begin{gather}
\ddot \rho+\frac{2}{r}\dot \rho -\frac{1}{2r^2} f^2 \rho
=-\frac14 (B-A)^2 \rho +\frac{\lambda}{2} \big(\rho^2
-\rho_0^2 \big) \rho, \nn \\
\ddot f - \frac{1}{r^2} (f^2 -1) f
=\big(\frac{g^2}{4} \rho^2-A^2 \big)~f,  \nn \\
\ddot A +\frac{2}{r} \dot A - \frac{2}{r^2} f^2 A
=\frac{g^2}{4}(A-B)~\rho^2,  \nn\\
\ddot B+\frac{2}{r}\dot B =\frac{g'^2}{4}(B-A)~^2,
\label{cmeq}
\end{gather}
which has a singular solution
\begin{gather}
f=0,~~~\rho=\rho_0 =\sqrt{2\mu^2/\lambda},  \nn\\
A_\mu^{\rm (em)} = -\frac{1}{e}(1-\cos \theta)
\pd_\mu \varphi,~~~Z_\mu=0.
\label{cmon}
\end{gather}
This describes the point monopole in Weinberg-Salam
model which has the magnetic charge $4\pi/e$
(not $2\pi/e$).

\begin{figure}
\includegraphics[height=4.5cm, width=8cm]{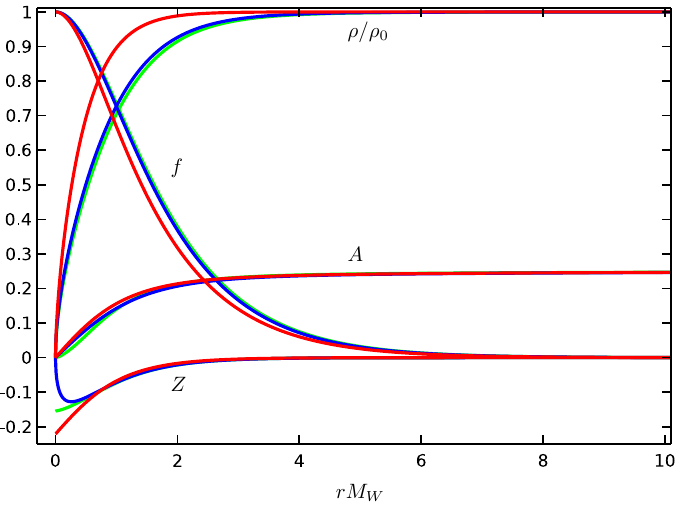}
\caption{\label{fed} The electroweak dyon solution.
The red and blue curves represent the singular Cho-Maison
dyon and the regularized finite energy dyon obtained by
(\ref{fedeq1}), and the green curves represent 
the regularized dyon obtained by (\ref{fedeq2})
with $n=6$. Notice that the blue and green curves are 
almost indistinguishable.}
\end{figure}

With the boundary condition
\begin{gather}
\rho(0)=0,~~f(0)=1,~~A(0)=0,~~B(0)=b_0, \nn\\
\rho(\infty)=\rho_0,~f(\infty)=0,
~A(\infty)=B(\infty)=A_0,
\label{bc0}
\end{gather}
we can integrate (\ref{cmeq}). With $A=B=0$ we have
the electroweak monopole with $q_m=4\pi/e$, but with
non-trivial $A$ and $B$ we have the electroweak dyon
which has the extra electric charge $q_e$ \cite{plb97,epjc15}.
The monopole and dyon solutions are shown in Fig. \ref{fem}
and Fig. \ref{fed} in red curves.

There have been many studies of the electroweak 
monopole \cite{ellis,ak,bb,mav,pta19,epjc20,fh}. In particular, the stability of the Cho-Maison monopole 
has been established \cite{gv1}, and multi-monopole 
solutions have been discovered \cite{gv2}. 

Since the solution contains the point singularity at
the origin, it can be viewed as a hybrid between the Dirac
monopole and the 'tHooft-Polyakov monopole. So at
the classical level it carries an infinite energy, 
so that the mass is not determined. However, we could 
regularize the point singularity with the quantum correction 
at short distance. We could do this replacing the $U(1)_Y$ 
coupling constant $g'$ to an effective coupling, 
or replacing the real electromagnetic coupling constant 
$e$ to an effective coupling \cite{epjc15}.

To show this, we modify the Lagrangian (\ref{lag1})
with a non-trivial $U(1)_Y$ permittivity
$\eps (\rho)$ which depends on $\rho$,
\begin{gather}
{\cal L}=-\frac{1}{2} (\partial_\mu \rho)^2
-\frac{\lambda}{8}\big(\rho^2 -\rho_0^2 \big)^2
- \frac{1}{4} {F'}_\mn^2  \nn\\
-\frac14  \eps(\rho)~G_\mn^2
-\frac{1}{2} |D'_\mu W_\nu-D'_\nu W_\mu|^2 \nn\\
- \frac{g^2}{4} {\rho}^2 W_\mu^* W_\mu
-\frac18 \rho^2 (gA'_\mu-g'B_\mu)^2  \nn\\
+ ig F'_\mn W_\mu^*W_\nu
+ \frac{g^2}{4}(W_\mu^* W_\nu - W_\nu^* W_\mu)^2.
\label{effl1}
\end{gather}
This retains the full $SU(2)\times U(1)_Y$ gauge symmetry.
Moreover, when $\eps$ approaches to one asymptotically,
it reproduces the standard model. But $\eps$ effectively
changes the $U(1)_Y$ gauge coupling $g'$ to the ``running"
coupling $\bg'=g' /\sqrt{\eps}$. So, by making $\bg'$
infinite at the origin, we can regularize the Cho-Maison
monopole \cite{epjc15}.

With this modification the dyon energy and the equation
of motion are given by
\begin{gather}
E=4\pi \int_0^\infty dr \bigg\{\frac{\eps}{2g'^2 r^2}
+\frac12 (r\dot\rho)^2
+\frac{\lambda}{8} r^2 \big(\rho^2-\rho_0^2 \big)^2 \nn\\
+\frac1{g^2} \big(\dot f^2 + \frac{(f^2-1)^2}{2r^2}
+ f^2 A^2 \big) +\frac14 f^2\rho^2  \nn\\
+\frac{(r\dot A)^2}{2g^2}+\frac{\eps (r\dot B)^2}{2g'^2}
+\frac{r^2}{8} (A-B)^2 \rho^2 \bigg\},
\label{fede1}
\end{gather}
and
\begin{gather}
\ddot \rho + \frac{2}{r}\dot \rho-\frac{f^2}{2r^2} \rho
=\frac{\lambda}{2} (\rho^2- \rho_0^2) \rho
-\frac{1}{4} (A-B)^2 \rho  \nn\\
+\frac{\eps'}{2 g'^2}\big(\frac{1}{r^4}-\dot{B}^2 \big),  \nn\\
\ddot{f}-\frac{f^2-1}{r^2}f
=\big(\frac{g^2}{4} \rho^2 - A^2\big)f, \nn\\
\ddot{A}+\frac{2}{r}\dot{A}-\frac{2f^2}{r^2}A
=\frac{g^2}{4} \rho^2(A-B), \nn \\
\ddot{B} + 2\big(\frac{1}{r}
+\frac{\eps'}{2 \eps} \dot \rho \big) \dot{B}
=-\frac{g'^2}{4 \epsilon} \rho^2 (A-B),
\label{fedeq1}
\end{gather}
where $\eps' = d\eps/d\rho$. Assuming
\begin{gather}
\rho(r) = r^\delta (h_0+h_1 r+...),   \nn\\
\eps =\Big(\frac{\rho}{\rho_0}\Big)^n
\big[c_0+c_1 (\frac{\rho}{\rho_0})+...\big],
\label{eps}
\end{gather}
near the origin we can show that the energy becomes
finite when $n > 2$ \cite{epjc15}.

We can integrate (\ref{fedeq1}) with
$\eps=(\rho/\rho_0)^n$. The regularized monopole
and dyon solutions with $n=6$ are shown in
Fig. \ref{fem} and Fig. \ref{fed} in blue curves.
Notice that asymptotically the regularized solutions
look very much like the singular solutions, except that
for the finite energy dyon solution $Z$ becomes zero
at the origin. The regularized monopole energy with $n=6$
becomes
\begin{gather}
E \simeq 0.65 \times \frac{4\pi}{e^2} M_W
\simeq 7.20 ~{\rm TeV}.
\end{gather}
This confirms that the ultraviolet regularization of
the Cho-Maison dyon is indeed possible.

We could also regularize the monopole with the real
electromagnetic permittivity. Consider the Lagrangian
\begin{gather}
{\cal L'}= -\frac12 (\pd_\mu \rho)^2
-\frac{\lambda}{8}\big(\rho^2-\rho_0^2 \big)^2
-\frac14 \beps(\rho) {F_\mn^{\rm (em)}}^2 \nn\\
-\frac12 \Big|\big(D_\mu^{\rm (em)}+ie\frac{g}{g'}Z_\mu\big) W_\nu
-\big(D_\nu^{\rm (em)}+ie\frac{g}{g'}Z_\nu\big) W_\mu\Big|^2  \nn\\
+ie (F_\mn^{\rm (em)} +\frac{g}{g'}  Z_\mn) W_\mu^* W_\nu
+ \frac{g^2}{4}(W_\mu^* W_\nu - W_\nu^* W_\mu)^2  \nn\\
-\frac14 Z_\mn^2 -\frac{g^2}{4}\rho^2 W_\mu^*W_\mu
-\frac{g^2+g'^2}{8} \rho^2 Z_\mu^2,
\label{effl2}
\end{gather}
where $\beps$ is the real non-vacuum electromagnetic
permittivity. Just like (\ref{effl1}) it retains all
symmetries of the standard model, and asymptotically
reduces to the standard model with $\beps \rightarrow 1$.
It has the dyon equation of motion
\begin{gather}
\ddot \rho + \frac{2}{r}\dot \rho-\frac{f^2}{2r^2} \rho
=\frac{\lambda}{2} (\rho^2- \rho_0^2) \rho
-\frac14 (B-A)^2 \rho    \nn\\
+\frac{\beps'}{2}\Big(\frac{1}{e^2 r^4}
-e^2(\frac{\dot A}{g^2}+\frac{\dot B}{g'^2})^2 \Big),  \nn\\
\ddot{f}-\frac{f^2-1}{r^2}f
=\big(\frac{g^2}{4} \rho^2 - B^2\big) f, \nn\\
\ddot A + \frac{2}{r} \dot A
+e^2 \frac{\beps'}{\beps}\dot \rho \big(\frac{\dot A}{g^2}
+\frac{\dot B}{g'^2}\big)
-\frac{2 e^2}{g^2} \big(\frac{g^2}{g'^2}
+\frac{1}{\beps}\big) \frac{f^2}{r^2} A  \nn\\
=-\frac{g^2}{4} \rho^2 (B-A),  \nn\\
\ddot B +\frac{2}{r}\dot B
+e^2 \frac{\beps'}{\beps} \dot \rho \big(\frac{\dot A}{g^2}
+\frac{\dot B}{g'^2}\big) +\frac{2e^2}{g^2} \big(1-
\frac{1}{\beps} \big) \frac{f^2}{r^2} A \nn\\
=\frac{g'^2}{4}(B-A) \rho^2.
\label{fedeq2}
\end{gather}
To integrate it out and find a finite energy solution
we have to choose a proper boundary condition.

Now, with
\begin{gather}
\beps =\eps_0 +\eps_1,   \nn\\
\eps_0= \frac{g'^2}{g^2+g'^2},
~~~\eps_1 =\eps_1(\rho),
\label{epbc}
\end{gather}
the dyon energy becomes
\begin{gather}
E=4\pi \int_0^\infty dr \bigg\{\frac{1}{2e^2 r^2}
\Big(\eps_0 (f^2-1)^2+ \eps_1 \Big)
+\frac12 (r\dot \rho)^2   \nn\\
+\frac{\lambda r^2}{8}\big(\rho^2-\rho_0^2 \big)^2
+\frac1{g^2} \dot f^2 +\frac14 f^2 \rho^2
+ \frac{f^2 A^2}{g^2}  \nn\\
+ \frac{r^2}{8}(A-B)^2\rho^2
+ \frac{r^2(A'-B')^2}{2(g^2+g'^2)}   \nn\\
+ (\eps_0+\eps_1) \frac{e^2 r^2}{2} \big(\frac{A'}{g^2}
+\frac{B'}{g'^2} \big)^2  \bigg\}.
\label{fede2}
\end{gather}
This shows that the energy can be made finite with $f(0)=1$,
when $\eps_1$ approaches to zero quickly enough near the origin.

We can integrate (\ref{fedeq2}) with
\begin{gather}
\beps =\eps_0 +(1-\eps_0) \Big(\frac{\rho}{\rho_0}\Big)^n.
\label{epbc}
\end{gather}
With $A=B=0$, we have the monopole solution
shown in Fig. \ref{fem} in red curves, and the dyon solution
shown in Fig. \ref{fed} with non-trivial $A$ and $B$ in blue
curves.

Remarkably, for the monopole solutions for the two equations
(\ref{fedeq1}) and (\ref{fedeq2}) become identical. To see 
this notice that with
\begin{gather}
\beps'=\frac{\pd}{\pd \rho}\Big[\eps_0
+(1-\eps_0)\big(\frac{\rho}{\rho_0}\big)^n \Big]
=\frac{g^2}{g^2+g'^2}\frac{\pd}{\pd \rho}
\Big(\frac{\rho}{\rho_0}\Big)^n,  \nn\\
\frac{\beps'}{e^2}=\frac{1}{g'^2}\frac{\pd}{\pd \rho} \Big(\frac{\rho}{\rho_0}\Big)^n
=\frac{1}{g'^2} \eps',
\end{gather}
the monopole equation of (\ref{fedeq2}) with $A=B=0$ becomes
\begin{gather}
\ddot \rho + \frac{2}{r}\dot \rho-\frac{f^2}{2r^2} \rho
=\frac{\lambda}{2} (\rho^2- \rho_0^2) \rho
+\frac{\eps'}{2g'^2r^4},  \nn\\
\ddot{f}-\frac{f^2-1}{r^2}f =\frac{g'^2}{4} \rho^2 f.
\label{}
\end{gather}
This is identical to the monopole equation (\ref{fedeq1})
with $A=B=0$. Moreover, with
\begin{gather}
\frac{1}{e^2}\eps_0=\frac{1}{e^2}\frac{g'^2}{g^2+g'^2}
=\frac{1}{g^2},\nn\\
\frac{1}{e^2}(1-\eps_0)=\frac{1}{e^2}\frac{g^2}{g^2+g'^2}
=\frac{1}{g'^2},
\end{gather}
the monopole energy (\ref{fede2}) with $A=B=0$
becomes identical to the monopole energy given by (\ref{fede1})
with $A=B=0$. This assures that the two monopole solutions
regularized by the hypercharge renormalization and the real electric charge renormalization are indeed identical to 
each other \cite{epjc20}. For the dyon, however, the two 
equations (\ref{fedeq1}) and (\ref{fedeq2}) give different solutions. This is evident in Fig. \ref{fed}.

Of course, the above regularization is not the only way 
to regularize the Cho-Maison monopole. One could regularize 
the monopole replacing the $U(1)_Y$ part by the Bonn-Infeld Lagrangian \cite{ak,mav}, or by other (logarithmic or 
exponential) non-linear extensions of the $U(1)_Y$ part 
in (\ref{lag1}) \cite{fh}. For example, in the Bonn-Infeld modification the monopole mass is estimated to be around 
11.6 TeV, and in the logarithmic and exponential extensions 
the mass is estimated to be around 8.7 TeV and 7.9 TeV. 

This, with the above analysis of the regularization 
by charge renormalization, strongly suggests that 
the monopole could also be regularized replacing 
the real Maxwell part (not the hypercharge part) by 
the Bonn-Infeld Lagrangian or by the nonlinear extensions 
in (\ref{lag2}).

\section{Electroweak String Configuration}

As we have pointed out, the standard model has two
$\pi_1(S^1)$ topology, so that it should have strings.
To see this notice that the core of the electroweak monopole
is the Dirac type singular monopole shown in (\ref{cmon}).
If this is so, we could also expect the singular electroweak
string made of monopole-antimonopole pair infinitely
separated apart, which can be expressed as
\begin{gather}
\rho=\rho_0,
~~~A_\mu^{\rm (em)} = \frac{2n}{e}~\pd_\mu \vp,   \nn\\
W_\mu=0,~~~Z_\mu=0,
\label{cstring}
\end{gather}
which carries the quantized magnetic flux $4\pi n/e$.
This strongly implies that the string made of the Cho-Maison monopole-antimonopole pair could exist.

To obtain such solution we consider the following string 
ansatz in the cylindrical coordinates $(t,r,\varphi,z)$, 
\begin{gather}
\phi= \frac{1}{\sqrt 2} \rho(r) \xi,
~~~\xi= \left(\begin{array}{cc}
-\sin \dfrac{\al(r)}{2}~\exp (-in \vp) \\
\cos \dfrac{\al(r)}{2} \end{array} \right), \nn\\
\A_\mu=\frac{n}{g} \big(A(r)+1-\cos\al(r) \big) \pd_\mu \vp~\n   \nn\\
+\frac1g \big(f(r)-1 \big)~\n \times \pd_\mu \n,  \nn\\
B_\mu = \frac{m}{g'} B~\pd_\mu \vp,
\label{sans0}
\end{gather} 
where $m$ and $n$ are integers. To understand the physical 
meaning of the ansatz we notice that, with the $U(1)_Y$ 
gauge transformation
\begin{gather} 
\xi \rightarrow \exp \big(-i(m-\frac{n}2)\vp \big) \xi \nn\\
= \exp (-im \vp) \left(\begin{array}{cc}
-\sin \dfrac{\al(r)}{2}~\exp (-i\dfrac{n}2 \vp) \\
\cos \dfrac{\al(r)}{2}~\exp (i\dfrac{n}2 \vp)
\end{array} \right),   \nn\\
B_\mu \rightarrow \frac{1}{g'}\Big(m(B-2) +n \Big)~\pd_\mu \vp,
\label{sgt}
\end{gather} 
the ansatz acquires the following form 
\begin{gather}
\phi= \frac{1}{\sqrt 2} \rho(r) \xi,   \nn\\
\xi= \exp(-im\vp)\left(\begin{array}{cc}
~-\sin \dfrac{\al(r)}{2}~\exp (-i\dfrac{n}{2} \vp) \\
\cos \dfrac{\al(r)}{2}~\exp (i\dfrac{n}{2} \vp)
\end{array} \right),  \nn\\
\A_\mu=\frac{n}{g} \big(A(r)+1-\cos\al(r) \big) \pd_\mu \vp~\n  \nn\\
+\frac1g \big(f(r)-1 \big)~\n \times \pd_\mu \n,  \nn\\
B_\mu =\frac{1}{g'}\Big(m(B-2) +n \Big)~\pd_\mu \vp.
\label{sans1}
\end{gather}
In this expression the meaning of the integers $m$ and $n$ 
becomes clear. They represent the winding numbers of 
the $\pi_1(S^1)$ topology of $U(1)_Y$ and $U(1)$ subgroup 
of $SU(2)$. 

Notice, however, that in the ansatz (\ref{sans1}) $\xi$ 
becomes single valued under the $2\pi$ rotation along 
the z-axis only when $n$ becomes even integers. This is 
because the gauge transformation (\ref{sgt}) becomes singular 
when $n$ becomes odd integers. This means that for $n$ 
to represent the $\pi_1(S_1)$ topology of the $U(1)$ subgroup 
of SU(2) it must be even integers. This point will become 
important later.   

With (\ref{sans0}) we have 
\begin{gather}
\n=-\xi^\dagger \vsig \xi
=\left(\begin{array}{ccc} \sin \al \cos n\vp \\
	\sin \al \sin n\vp \\ \cos \al
\end{array} \right),  \nn\\
C_\mu \simeq -\frac{2i}{g} \xi^\dagger \pd_\mu \xi 
=-\frac{n}{g} (1-\cos \al)~\pd_\mu \vp,   \nn\\
A'_\mu= \frac{n}{g} A~\pd_\mu \vp.
\end{gather}
This means that the ansatz, in terms of the physical field, 
becomes
\begin{gather}
\rho=\rho(r),  \nn\\
A_\mu^{(\rm em)}= e \Big(\frac{n}{g^2} A
+\frac{m}{g'^2} B \Big)~\pd_\mu \vp,    \nn\\
W_\mu= \frac{i}{g} \frac{f}{\sqrt 2} \exp (in\vp)
~(\pd_\mu \al +i n \sin \al  \pd_\mu \vp),  \nn\\
Z_\mu=\frac{1}{\sqrt{g^2+g'^2}} 
\big(n A -m B \big)~\pd_\mu \vp.
\label{sans2}
\end{gather}
We can confirm this writing the ansatz in the ``unitary" 
(or ``physical") gauge with the following gauge 
transformation 
\begin{gather}
\xi \rightarrow U \xi = \left(\begin{array}{cc} 0 \\
1 \end{array} \right), \nn\\
U= \left(\begin{array}{cc} \cos \dfrac{\al}{2}, 
& \sin \dfrac{\al}{2} \exp (-in \vp) \\
-\sin \dfrac{\al}{2} \exp (in \vp), 
& \cos \dfrac{\al}{2} \end{array} \right). 
\end{gather}
With this gauge transformation the ansatz changes to
\begin{gather}
\A_\mu \rightarrow \frac{1}{g} \left(\begin{array}{ccc}
-f(n\cos(n\varphi) \sin \alpha~\pd_\mu \vp +\sin(n\varphi)\pd_\mu\alpha)\\
-f(n\sin(n\varphi) \sin \alpha~\pd_\mu \vp -\cos(n\varphi)\pd_\mu\alpha) \\ 
n A~\pd_\mu \vp \end{array} \right),  \nn\\
B_\mu \rightarrow \frac{m}{g'} B~\pd_\mu \vp.	
\end{gather}
From this we reproduce (\ref{sans2}).

So, when
\begin{gather}
n A= m B
\label{emsans}
\end{gather}
the ansatz describes the electromagnetic string
\begin{gather}
A_\mu^{(\rm em)}= \frac{n}{e} A~\pd_\mu \vp,   \nn\\
Z_\mu=0.
\label{ems}
\end{gather}
But when
\begin{gather}
ng'^2 A +mg^2 B =0,
\label{zsans}
\end{gather}
the ansatz describes the Z boson string
\begin{gather}
A_\mu^{(\rm em)}=  0, \nn\\
Z_\mu=-m \frac{\sqrt{g^2+g'^2}}{g'^2} B~\pd_\mu \vp 
=-\frac{m}{e} B~\cot~\theta_{\rm w}
~\pd_\mu \vp.
\label{zs}
\end{gather}
This confirms that the ansatz is able to describe both 
electromagnetic and Z boson strings. 

With the ansatz (\ref{sans0}) we have the following
string equations of motion from (\ref{em0}),
\begin{widetext}
\begin{gather}
\ddot \rho+\frac{\dot \rho }{r}
-\frac{1}{4} \Big(f^2 \dot \al^2 
+\frac{n^2 f^2 \sin^2 \alpha +(nA-mB)^2}{r^2} \Big)~\rho
=\frac{\lambda}{2}\rho(\rho^2-\rho_0^2),   \nn\\
f\rho\Big[\ddot{\alpha}+\Big(\frac{1}{r}
+2\frac{\dot{\rho}}{\rho}
+\frac{\dot{f}}{f}\Big)\dot{\alpha}
-\frac{n(mB+n)\sin\alpha}{r^2}\Big]=0,   \nn\\	
n \Big(\ddot A-\frac{\dot A}{r} \Big)
+n f^2 \Big(\sin \alpha (\ddot \alpha 
-\frac{\dot \alpha}{r}
+3 \frac{\dot f}{f} \dot \alpha)
+(2\cos \alpha -A-1) \dot \alpha^2 \Big)  
= \frac{g^2}{4} \rho^2 \big(nA-mB \big),   \nn\\
\sin \al \Big(\ddot f -\frac{\dot{f}}{r} 
-(f^2+1) f \dot \al^2 \Big)
+(\cos \al-A-1)f \Big(\ddot \al -\frac{\dot{\al}}{r}\Big) 
+(2\cos \al-A-1) \dot \al \dot f -2 f \dot A \dot \al   =\frac{g^2}{4} \rho^2f \sin \al,  \nn\\		
m \big(\ddot B-\frac{\dot B}{r} \big)
= -\frac{g'^2}{4} \rho^2 \big(nA-mB \big),  \nn\\
n^2 \Big[(A+1) \sin \al \dot f -\Big(f^2 \sin^2 \al
-(A+1)(\cos \al-A-1)\Big) f \dot \al
-f \sin \al \dot A \Big]=\frac{g^2}{4} r^2\rho^2 f \dot \al.
\label{seq}	
\end{gather}
\end{widetext}
Notice that, in spite of the fact that we have 5 variables
in the ansatz (\ref{sans0}), here we have 6 equations of 
motion. So in principle, we could have no solution. As we 
will see, however, the equation does allow us to have 
solutions.  

\section{Electromagnetic String: Quantized Magnetic Vortex}

From (\ref{seq}) we can obtain the electromagnetic 
string solution made of quantized magnetic flux. 
With (\ref{emsans}) the equation reduces to
\begin{gather}
\ddot \rho+\frac{\dot \rho}{r}
-\frac{f^2}{4} \Big(\dot \al^2 
+\frac{n^2 \sin^2 \al}{r^2} \Big)~\rho
=\frac{\lambda}{2}\rho(\rho^2-\rho_0^2),   \nn\\
f\rho\Big[\ddot \al+\Big(\frac{1}{r}+2\frac{\dot{\rho}}{\rho}
+\frac{\dot{f}}{f}\Big)\dot \al
-\frac{n^2\sin \al (A+1)}{r^2}\Big]=0,   \nn\\	
f^2 \Big(\sin \al (\ddot \al -\frac{\dot \al}{r}
+3 \frac{\dot f}{f} \dot \al)
+(2\cos \al -A-1) \dot \al^2 \Big) = 0,   \nn\\
\sin \al \Big(\ddot f -\frac{\dot{f}}{r} 
-(f^2+1) f \dot \al^2 \Big) 
+(\cos \al -A-1)f \Big(\ddot \al -\frac{\dot{\al}}{r}\Big) \nn\\  
+(2\cos \al -A-1) \dot \al \dot f -2 f \dot A \dot \al 
=\frac{g^2}{4} \rho^2 f \sin \al,  		\nn\\
\ddot A-\frac{\dot A}{r} = 0,  \nn\\
n^2 \Big[\sin \al \Big((A+1) \dot f -\dot A f \Big)  \nn\\
-\Big(f^2 \sin^2 \al 
-(A+1)(\cos \al-A-1) \Big)f \dot \al \Big]   \nn\\
=\frac14 g^2r^2\rho^2 f \dot \al.
\label{emseq}
\end{gather}
This, with 
\begin{gather}
A=-1,~~~\al=\frac{\pi}{2}
\label{emscon}
\end{gather}
reduces to
\begin{gather}
\ddot \rho+\frac{\dot \rho}{r}
-\frac{n^2}{4} \frac{f^2}{r^2}~\rho
=\frac{\lambda}{2}\rho(\rho^2-\rho_0^2),   \nn\\	
\ddot f -\frac{\dot f}{r} -\frac{g^2}4 \rho^2 f =0.
\label{emseq1}
\end{gather}	
Notice that this is precisely the equation which describes 
the Abrikosov-Nielsen-Olesen (ANO) vortex \cite{ano}.
But here we are not looking for the ANO vortex, 
but an electromagnetic vortex described by (\ref{ems}).

Clearly (\ref{emseq1}) has the naked singular electromagnetic
string solution given by
\begin{gather}
\rho=\rho_0,~~~~~f=0,    \nn\\
A_\mu^{(\rm em)}= -\frac{n}e \pd_\mu \vp,
\label{sems}
\end{gather}
which becomes precisely the magnetic vortex solution 
that we predicted in (\ref{cstring}). The only difference 
is that here the string carries the quantized magnetic flux  
$2\pi n/e$, not $4\pi n/e$. So only when $n$ becomes even,
the above solution describes the predicted solution.

To understand the situation we have to remember that 
only when $n$ becomes even integers it represents 
the $\pi_1(S^1)$ topology of the $U(1)$ subgroup of 
$SU(2)$. This is because the period of the $U(1)$ 
subgroup of $SU(2)$ is $4\pi$, not $2\pi$. So the above 
solution indeed describes the string solution made of 
the monopole-antimonopole pair infinitely separated 
apart with the $\pi_1(S^1)$ topology of the $U(1)$ 
subgroup of $SU(2)$, when $n$ becomes even. But what is 
interesting here is that we have the electromagnetic 
string even when $n$ is odd, which can not be interpreted 
as the string made by monopole-antimonopole pair.   

\begin{figure}
\includegraphics[height=4.5cm, width=8cm]{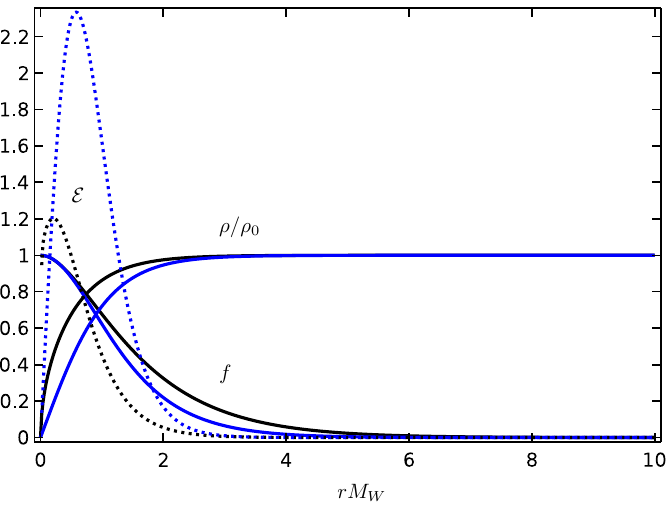}
\caption{\label{mvs1} The Higgs and W boson configurations
of the quantized magnetic vortex solutions with $A=-1$ and
$\al=\pi/2$. The black and blue curves represent solutions
with $n=1$ and $n=2$, respectively. The dotted curves
represent the energy densities of the Higgs and W bosons.}
\end{figure}

We can solve (\ref{emseq1}) with the boundary condition
\begin{gather}
\rho(0)=0,~~~\rho(\infty)=\rho_0,   \nn\\
f(0)=1,~~~f(\infty)=0,
\end{gather}
and obtain the singular but dressed quantized string solution shown in Fig. (\ref{mvs1}), dressed by Higgs and W boson. At the origin 
$\rho$ and $f$ can be expressed by
\begin{gather}
\rho\simeq r^\delta(a_1+a_2 r+...),~~~\delta=|n|/2, \nn\\
f\simeq 1+  b_1 r^2+...,
\end{gather}
so that the solution becomes regular. 

Asymptotically they have the form of modified second 
kind Bessel function
\begin{gather}
\rho=\rho_0-K_0(\sqrt{\lambda\rho_0}r)\nn\\
\simeq\rho_0 -\sqrt{\frac{\pi}{2\sqrt{\lambda\rho_0}r}}
\exp(-\sqrt{\lambda\rho_0}r)\left(1
-\frac{1}{8\sqrt{\lambda\rho_0}r}+...\right),\nn\\
f=rK_1\left(\frac{g\rho_0}{2}r\right)\nn\\
\simeq\sqrt{\frac{\pi r}{g\rho_0}}\exp\left(-\frac{g\rho_0}{2}r\right)
\left(1+\frac{3}{4g\rho_0r}+...\right)
\end{gather}
so that they have the exponential damping set by the Higgs
and W boson mass
\begin{gather}
\rho \simeq \rho_0 -\sqrt{\frac{\pi}{2 m_H r}} \exp (-m_H r)+...,   \nn\\
f \simeq \sqrt{\frac{\pi r}{2m_W}} \exp (-m_W r)+...,
\label{dems}
\end{gather}
similar to the Cho-Maison monopole \cite{plb97}. This 
confirms that the dressed string solution is nothing but 
the string made of the Cho-Maison monopole-antimonopole 
pair infinitely separated apart.  

To understand the meaning of this solution, remember that
the electromagnetic field of the solution is given by
\begin{gather}
A_\mu^{(\rm em)}= -\frac{n}{e}\pd_\mu \vp,
\label{emv}
\end{gather}
which has the quantized magnetic flux along the $z$-axis,
\begin{gather}
\Phi=\Int \B \cdot d\vec S
=\Int_{r=\infty} A_\mu^{\rm (em)} dx^\mu
= -\frac{2\pi n}{e}.
\end{gather}
So the solution describes the singular quantized magnetic
flux dressed by the Higgs and W bosons in the standard
model. Notice that, although the magnetic field $\B$ vanishes
everywhere except at the origin, it has a 2-dimensional $\delta$-function singularity at the origin. Clearly this 
string singularity is topological, whose quantized magnetic 
flux represents the non-trivial winding number of $\pi_1(S^1)$. 
This means that, just like the Cho-Maison monopole, the solution has infinite magnetic energy (per unit length). 

For the monopole, it is well known that the magnetic 
point singularity which makes the energy infinite can 
be regularized by various methods, for example by 
the vacuum polarization or by the gravitational 
interaction \cite{epjc15,bb,pta19,epjc20}. One might 
wonder if similar methods could also make the string 
energy finite. This is a very interesting question 
to be studied further.

\begin{figure}
\includegraphics[height=4.5cm, width=8cm]{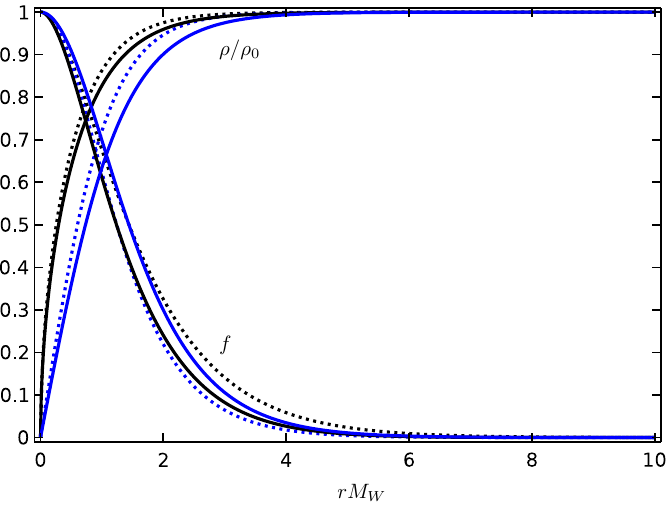}
\caption{\label{bpss} The BPS string solutions
with $A=-1$ and $\al=\pi/2$. The black and blue curve
represent solutions with $n=1$ and $n=2$, respectively.
For comparison the singular string solutions in
Fig. \ref{mvs1} are plotted correspondingly with dotted
black lines and blue lines.}
\end{figure}

Excluding the singular magnetic field, the energy
(per unit length) of the Higgs and W boson is given by
\begin{gather}
E =2\pi \Int \Big[\frac{\dot\rho^2}{2}
+\frac{\lambda}{8}(\rho^2-\rho_0^2)^2
+\frac{n^2\dot{f}^2}{2g^2r^2}   \nn\\
+\frac{n^2f^2\rho^2}{8r^2} \Big]rdr.
\end{gather}
This is the same energy functional of Abrikosov-Nielsen-Olesen
vortex which has the following BPS bound,
\begin{gather}
E = 2\pi \Int \Big[\frac{1}{2g^2r}\Big(n\dot f
+\frac{g\sqrt{\lambda}}{2}r(\rho^2-\rho^2_0) \Big)^2\nn\\
+\frac{r}{2}(\dot\rho-\frac{nf\rho}{2r})^2
-\frac{n\sqrt{\lambda}}{2g}\dot f(\rho^2-\rho^2_0)
+\frac{1}{2}nf\rho\dot\rho \Big]dr \nn\\
\geq 2\pi \Int \Big[ \frac{\lambda}{4}r(\rho^2-\rho_0^2)^2
+\frac{n^2f^2\rho^2}{4r}\Big]dr,
\end{gather}
when we have the Bogomolny equation
\begin{gather}
\dot\rho=\frac{nf\rho}{2r},\nn\\
n\dot f=-\frac{g\sqrt\lambda}{2}r(\rho^2-\rho_0^2).
\end{gather}
In the BPS limit (with $\lambda=0$) this has the trivial
solution (\ref{sems}). For $\lambda \neq 0$ the numerical
BPS vortex solution is shown in Fig. \ref{bpss}.

As we have already pointed out, (\ref{ems}) is 
the equation for the ANO vortex. So, as far as $\rho$ 
and $f$ are concerned, they describe the well known 
ANO vortex. The difference here is that we also have 
the singular quantized magnetic vortex (\ref{emv}). 
If so, one might wonder what will happen if we remove
the string singularity of (\ref{emv}) with a gauge 
transformation.

To answer this we consider the gauge transformation
\begin{gather}
A_\mu^{(\rm em)} \rightarrow {A'}_\mu^{(\rm em)}
=A_\mu^{\rm (em)} +\frac{n}{e} \pd_\mu \vp =0,   \nn\\
W_\mu \rightarrow W_\mu'
=\exp \big(-in \vp \big) W_\mu
= -\frac{n}{g} \frac{f}{\sqrt 2} \pd_\mu \vp.
\label{sgtw}
\end{gather}
This is the singular gauge transformation which removes 
the string singularity and changes the $\pi_1(S^1)$ 
topology of the string. Obviously this changes the physical 
content of the solution. But mathematically there is nothing 
wrong with this gauge transformation, in the sense that it 
keeps ${A'}_\mu^{(\rm em)}$ and $W_\mu'$ as a qualified 
solution after the gauge transformation. So, after the gauge 
they become physically a different solution. This tells 
that the standard model has a electromagnetically neutral 
string solution made of Higgs and W bosons described by 
Fig. \ref{mvs1} which does not carry any magnetic flux, 
whose energy is finite. 

The reason why such a solution is possible is that, 
in the absence of $A_\mu^{\rm (em)}$ and $Z_\mu$,
the Lagrangian (\ref{lag2}) reduces to the Landau-Ginzburg
theory with the gauge potential $W'_\mu$ when $W_\mu$ 
becomes $W'_\mu$. So it must have the ANO vortex solution, 
which is exactly the solution discussed above. In fact we 
can easily see that the equation (\ref{emseq1}) is exactly 
the equation for the ANO vortex. In the literature this 
solution is known as the W string \cite{vacha,bvb}. This 
tells that the electromagnetic vortex solution in the standard 
model is nothing but the W string which has the topological 
singular quantized magnetic flux at the core. 

Notice that to have the above quantized magnetic vortex 
we do not have to require $\al$ to be $\pi/2$. Actually 
$\al$ can be any constant, as far as $\sin \al \neq 0$. 
In fact, with $A=-1$ and $\sin \al \neq 0$, (\ref{emseq}) 
reduces to 
\begin{gather}
\ddot \rho+\frac{\dot \rho }{r}
-\frac{n^2}4 \frac{f^2}{r^2} \sin^2 \alpha~\rho
=\frac{\lambda}{2}\rho(\rho^2-\rho_0^2),   \nn\\
\sin \al \Big(\ddot f -\frac{1}{r} \dot f
-\frac{g^2}{4} \rho^2 f \Big)=0.
\label{emseq2}
\end{gather}
Obviously this (with the replacement of $f \sin \al$ to $f$)
is mathematically identical to (\ref{emseq1}), and has
essentially the same quantized magnetic vortex solutions.
 
\section{Z string}

To obtain the Z string solution, notice that 
(\ref{zsans}) reduces the string equation 
(\ref{seq}) to 
\begin{gather}
\ddot \rho+\frac{\dot \rho }{r}
-\frac14 \Big(f^2 \dot \al^2 +n^2 \frac{f^2 \sin^2 \al
+A^2\sec^4\theta_W}{r^2} \Big)~\rho  \nn\\
=\frac{\lambda}{2}\rho(\rho^2-\rho_0^2),   \nn\\
f\rho\Big[\ddot \al +\Big(\frac{1}{r}+2\frac{\dot{\rho}}{\rho}
+\frac{\dot{f}}{f} \Big) \dot \al
-n^2\frac{(1-A\tan^2\theta_W) \sin \al}{r^2}\Big]=0,   \nn\\	
\Big(\ddot A-\frac{\dot A}{r} \Big)
+ f^2 \Big(\sin \al (\ddot \al -\frac{\dot \al}{r}
+3 \frac{\dot f}{f} \dot \al) \nn\\
+(2\cos \al -A-1) \dot \al^2 \Big) 
=\frac{g'^2+g^2}{4} \rho^2 A,   \nn\\
\sin \al \Big(\ddot f -\frac{\dot{f}}{r} 
-(f^2+1) f \dot \al^2 \Big)
+(\cos \al-A-1)f \Big(\ddot \al -\frac{\dot \al}{r}\Big) \nn\\
+(2\cos \al -A-1) \dot \al \dot f -2 f \dot A \dot \al   =\frac{g^2}{4} \rho^2f \sin \al,  \nn\\		
\big(\ddot A-\frac{\dot A}{r} \big)
=\frac{g'^2+g^2}{4} \rho^2 A,  \nn\\
n^2 \Big[(A+1) \sin \al \dot f -f \sin \al \dot A \nn\\
-\Big(f^2 \sin^2 \al-(A+1)(\cos \al-A-1)\Big) 
f \dot \al \Big]  \nn\\
=\frac14 g^2r^2\rho^2 f \dot \al.
\label{seq2}
\end{gather}	
where $\theta_W$ is the Weinberg angle.
 
When $f=0$, this simplifies to
\begin{gather}
\ddot \rho+\frac{\dot \rho}{r}
-n^2 \frac{Z^2}{r^2} \rho 
=\frac{\lambda}{2}\rho(\rho^2-\rho_0^2),   \nn\\
\ddot Z-\frac{\dot Z}{r} -\frac{g^2}{4}\rho^2 Z =0.
\label{zseq}
\end{gather}
where 
\begin{gather}
Z(r) =\frac{g^2+g'^2}{2 g'^2} A(r)
=-\frac{m}{2n} \frac{\cos^2 \theta_W}{\sin^4 \theta_W}~B(r).
\end{gather}
Clearly this is mathematically identical to the equation 
(\ref{emseq1}) which describes the well known ANO 
vortex \cite{ano}. The reason for this is that
the standard model reduces to Landau-Ginzburg theory 
in the absence of the $W$ boson and electromagnetic 
field, which is well known to admit the ANO vortex solution. 
This means that the standard model has another string 
solution known as the Z string in the literature \cite{vacha,bvb}. 

\begin{figure}
\includegraphics[height=4.5cm, width=8cm]{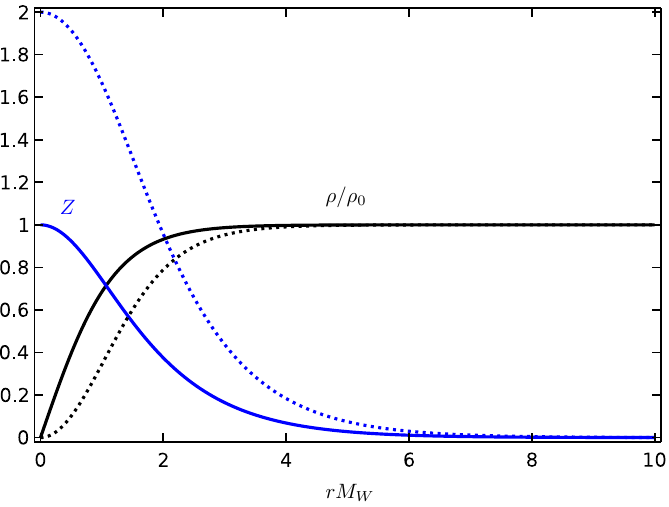}
\caption{\label{zstring} The Z boson string solution
for $k=1$ (solid curves) and $k=2$ (dotted curves)
with $n=1$.}
\end{figure}

We can solve (\ref{zseq}) with the boundary condition 
\begin{gather}
\rho(0)=0,~~~\rho(\infty)=\rho_0,   \nn\\
Z(0)=1,~~~Z(\infty)=0.
\end{gather} 
But here we can choose a more general boundary condition
\begin{gather}
\rho(0)=\frac{d\rho}{dr}(0)=...
=\frac{d^{k-1}\rho}{d^{k-1}r}(0)=0, \nn\\
\frac{d^k \rho}{d^k r}(0)\neq 0,   
~~~\rho(\infty)=\rho_0,  \nn\\
Z(0)=\pm\frac{k}{n},~~~Z(\infty)=0,
\end{gather}
and integrate (\ref{zseq}) to obtain the Z string 
solution given by
\begin{gather}	
\rho=\rho(r),   \nn\\
A_\mu^{(\rm em)}= 0,~~~W_\mu=0,  \nn\\
Z_\mu=- \frac{m}{e} B(r)~\cot \theta_W  \pd_\mu \vp.
\label{zst}
\end{gather}
The solution for $k=1$ and $k=2$ (with $n=1$) are 
shown in Fig. \ref{zstring}. This, of course, is 
the well known Z string \cite{vacha,bvb}. This confirms 
that the Z string is nothing but the ANO string 
embedded in the standard model.

Before we close this section it is worth mentioning 
another string in the standard model known as the Nambu 
string \cite{nambu}. To understand the Nambu string we 
notice that, with $\phi=(0,\eta)$, $\A_\mu=(0,0,A_\mu)$, 
and $B_\mu=0$, the Weinberg-Salam Lagrangian (\ref{lag0}) 
reduces to
\begin{gather}
{\cal L} =-|D_\mu \eta|^2
-\frac{\lambda}{2}\big(|\eta|^2
-\frac{\mu^2}{\lambda}\big)^2-\frac{1}{4} F_\mn^2, \nn \\
D_\mu \eta =\big(\pd_\mu +i\frac{g}{2} A_\mu \big) \eta.
\label{u1lag}
\end{gather} 
This is nothing but the Landau-Ginzburg Lagrangian which 
admits the ANO vortex, which carries the quantized magnetic 
flux $4\pi n/g$ (or $4\pi n \sin \theta_{\rm w} /e$)
of the gauge filed $A_\mu$. This vortex is known as 
the Nambu string. But physically this vortex is a mixed 
string made of the electromagnetic and Z strings, because 
$A_\mu$ is a combination of $A_\mu^{\rm (em)}$ and $Z_\mu$. 
So this string carries fractional (electro)magnetic and Z 
flux given by $(4\pi n/e) \times \sin^2 \theta_{\rm w}$ and
$(4\pi n/e) \times \sin \theta_{\rm w} \cos \theta_{\rm w}$. 
This is because $A_\mu$ is given by 
$A_\mu^{\rm (em)} \sin \theta_{\rm w}$ and 
$Z_\mu \cos \theta_{\rm w}$.

At this point one might worry about the stability of 
the above strings. This is ligitemate because, if they 
are not stable, they can not be treated as real. 
Fortunately we do not have to worry about this, since 
there is a simple and natural way to make them stable. 
Indeed we can always make them topologically stable 
by making them a twisted vortex ring, making them periodic 
and join both ends together, endowing the knot topology 
$\pi_3(S^2)$ \cite{prl01,prb06}. 

In fact it is well known that the Faddeev-Niemi knot 
in Skyrme theory can be interpreted as the twisted 
vortex ring (i.e., the twisted baby skyrmion) made 
this way. Of course, this type of twisted vortex ring
is, strictly speaking, not a string. But in all practical purpose they can be treated as a string, as far as we 
make them long enough. This assures us to tell confidently that the standard model does have the above Nambu
string.  
 
\section{Discussion}

In this work we have discussed all possible string type 
solutions in the standard model. We have shown that 
there are basically two types of topological string 
solutions in the standard model. The reason for this 
two types of solutions comes from the fact that 
the Weinberg-Salam Lagrangian contains two $U(1)$, 
the $U(1)$ subgroup of $SU(2)$ and $U(1)_Y$, which allows 
two different $\pi_1(S^1)$ string topology. 

The string solutions we have discussed include the well known Nambu string, Z string, and W string \cite{nambu,vacha,bvb}. 
The new solution here is the electromagnetic string 
which has the quantized $4\pi n/e$ magnetic flux which 
connects the Cho-Maison monopole-antimonopole pair 
infinitely separated apart. Actually there are two 
of them, a naked magnetic flux string and a dressed 
one which has the Higgs and W boson profiles. The existence 
of such quantized magnetic flux strings in the standard 
model, of course, is not surprising because the Cho-Maison monopole predicts this.   

In this connection it should be mentioned that a finite 
length quantized magnetic flux made of the Cho-Maison 
monopole-antimonopole pair has recently been constructed 
numerically \cite{wong}. It has the pole separation 
length around $8.2 / M_W$ and has the magnetic dipole 
moment roughly $8.6/M_W$. This is very interesting. 
The main difference between this and our solution is 
that this numerical string has a finite length, so that 
become metastable. This is because the monopole-antimonopole 
pair could annihilate each other. In comparison, ours 
is classically stable because it has infinite length. 
On the other hand the existence of the metastable solution 
made of Cho-Maison monopole-antimonopole pair strongly 
supports the existence of our string. With this we may 
conclude that the electromagnetic string made of 
quantized magnetic flux must exist in the standard model.     

The Cho-Maison monopole in the standard model has 
deep implications in physics, in particular, in 
cosmology. When coupled to gravity, the electroweak 
monopole turns to Reissner-Nordstrom type magnetic 
black hole which could become the seed of the stellar 
objects and galaxies. Moreover, they could account for 
the dark matter of the universe and the intergalactic 
magnetic field, and become the source of the ultra 
high energy cosmic rays \cite{pta19}. In fact we believe 
that the recently observed ultra high energy cosmic 
ray by the Telescope Array detector could have been 
generated by one of the remnant relativistic Cho-Maison monopoles created in the early universe \cite{ta}.  

Similarly we hope that the above strings in the standard 
model, in particular the electromagnetic quantized 
magnetic flux string, could play important roles in 
the formation of the large scale structure of the universe
in cosmology, as Witten has suggested \cite{witt}. 
For example, when coupled to gravity, our quantized 
magnetic flux string could become a black string (a (2+1)-dimensional black hole) which could play important 
roles in cosmology. This is a very interesting possibility 
worth to be studied further.

It must be emphasized that the importance of 
the electroweak strings comes from the fact that 
they exist in the standard model, so that we cannot 
ignore it. We must face it and deal with it, as long 
as the standard model is correct. We hope that our 
work in this paper could help us to understand 
the electroweak strings better.

{\bf Acknowledgments}

~~~The work is supported in part by the National Research
Foundation of Korea funded by the Ministry of Education
(Grant 2018-R1D1A1B0-7045163) and Ministry of Science and Technology (Grant 2022-R1A2C1006999), National Natural 
Science Foundation of China (Grant 12175320 and Grant 12375084) and National Science Foundation of Guangdong Province 
of China (grant 2022A1515010280), and by Center 
for Quantum Spacetime, Sogang University, Korea.

\end{document}